\documentstyle[prl,aps,epsfig]{revtex}

\newcommand{\beq}{\begin{equation}}
\newcommand{\eeq}{\end{equation}}
\newcommand{\beqar}{\begin{eqnarray}}
\newcommand{\eeqar}{\end{eqnarray}}
\newcommand{\bfig}{\begin{figure}}
\newcommand{\efig}{\end{figure}}

\newcommand{\bd}{\begin{itemize}} 
\newcommand{\ed}{\end{itemize}} 
\newcommand{\bc}{\begin{center}}
\newcommand{\ec}{\end{center}}
\newcommand{\be}{\begin{equation}}
\newcommand{\ee}{\end{equation}}

\newcommand{\ba}{\begin{array}}
\newcommand{\ea}{\end{array}}

\newcommand{\set}[2]{\newcommand{#1}{#2}}
\set{\pa}{\partial \over \partial\, }
\set{\leftvector}{\stackrel{\leftarrow}{\partial }}
\set{\rightvector}{\stackrel{\rightarrow}{\partial }}
\begin{document}
\twocolumn[\hsize\textwidth\columnwidth\hsize
           \csname @twocolumnfalse\endcsname

\title{Charged-current neutrino-$^{208}Pb$ reactions}
\author{C. Volpe$^{a,b}$, N.Auerbach$^{c}$, G.Col\`o$^{d}$, N. Van Giai$^{a}$}
\address{$^{a)}$
  Groupe de Physique Th\'{e}orique, Institut de Physique Nucl\'{e}aire,
F-91406 Orsay Cedex, France\\
$^{b)}$Institut f\"ur Theoretische Physik der Universit\"at, Philosophenweg
19, D-69120 Heidelberg, Germany \\
$^{c)}$
School of Physics and Astronomy, Tel Aviv University, Tel Aviv 69978,
Israel \\
$^{d)}$
Dipartimento di Fisica, Universit\`a degli Studi, and INFN, via Celoria 16, 
20133 Milano, Italia\\ }

\maketitle

\begin{abstract}
We present theoretical results
on the non flux-averaged $^{208}Pb(\nu_{e},e^-)^{208}Bi$ and 
$^{208}Pb(\nu_{\mu},\mu^-)^{208}Bi$
reaction cross sections,
obtained within the charge-exchange 
Random-Phase-Approximation. 
A detailed knowledge of these
cross sections is important in different contexts. In particular,
it is necessary to assess the possibility of using
lead as a detector in future experiments on supernova neutrinos,
such as OMNIS and LAND, and 
eventually detect neutrino oscillation signals by exploiting 
the spectroscopic properties of $^{208}Bi$.  
We discuss the present status on 
the theoretical predictions of
the reaction cross sections.
\end{abstract}

\vskip2pc
]

\noindent

The study of reactions induced by neutrinos on nuclei
is at present an active field of research. A detailed knowledge
of the reaction cross sections is interesting for different domains,
going from high energy physics to astrophysics \cite{ORLAND}.
For example, they
are  necessary in the interpretation of current experiments
on neutrinos as well as in the evaluation of possible new detectors
for future experiments. The importance of neutrino-nuclei reactions
in astrophysical processes, such as
the r-process nucleosynthesis, is also being attentively 
studied \cite{qian,Bor}.
In particular, $\nu-Pb$ reactions have attracted much interest
recently. Lead has been used as a shielding material
in the recent experiments on neutrino oscillations performed
by the LSND collaboration \cite{LSND1,LSND2} so that estimates of 
the $\nu-Pb$ reaction cross sections are necessary for the evaluation
of backgrounds in these experiments; also projects on
lead-based detectors \cite{LeadPer},
such as OMNIS \cite{Cline,OMNIS} and LAND \cite{LAND}, are
being studied for the purpose of
detecting supernova neutrinos. These detectors
might provide information on neutrino properties, such as oscillations
in matter \cite{Hax} or the mass \cite{Boyd1} by measuring the time delay
and/or spreading in the neutrino signal \cite{OMNIS,LAND} as well as
help in testing supernova models.
From the practical point of view,
lead-based detectors seem to 
present several of the characteristics required
to be supernova observatories, namely
high sensitivity to neutrinos of all flavors,
simplicity, reliability with inexpensive materials
\cite{LAND}.
Large cross sections for neutrinos
in the supernova energy range are also
an important condition since they determine the possible
rates and therefore the maximum observable distance.
Actually, $\nu$-nucleus reaction cross
sections increase strongly with the charge  of the nucleus.
For example, if the neutrinos come from the
Decay-At-Rest (DAR) of $\mu^+$, the cross sections of
the flux-averaged charged-current (CC) reaction
$\nu_e + _{Z}X_{N} \rightarrow _{Z+1}X'_{N-1} + e^-$ goes
from  about $14.~10^{-42}$ cm$^2$ for $^{12}C$ 
\cite{LSNDve,Allen,KAR_ve_Ngs},
to $2.56~10^{-40}$ cm$^2$ in $^{56}Fe$ \cite{Lan00}
and is estimated to be $3.62~10^{-39}$ in $^{208}Pb$ \cite{Lan00}.
Besides these practical features which are essential
in the choice of the nucleus to use to detect neutrinos,
another important feature is the 
spectroscopic properties
which may
suggest attractive signals of supernova neutrino oscillations.
In \cite{Hax}, for example, it has been shown that the measurement
of events where two neutrons are emitted by $^{208}Bi$ excited
in the reaction
$\nu_e + ^{208}Pb \rightarrow ^{208}Bi + e^-$
is both flavor-specific and very sensitive to the mean energy
of the $\nu_e$.
In case when
$\nu_{\mu},\nu_{\tau} \rightarrow \nu_e$ oscillations take place,
the hotter $\nu_e$  would increase the number of
two neutron events by a factor of forty \cite{Hax}.
Another possible signal has been proposed in \cite{Lan00}, that is that
the energy distribution of the neutrons emitted in the same CC
reaction  should have a peak at low energy
more or less pronounced according to whether
the oscillations occur or not.
This peak would come from the excitation of
a peak at around 8 MeV in the Gamow-Teller strength
distribution. (One should however note that
this peak has never been observed experimentally).
Both the estimate
of the CC $\nu-Pb$ reaction cross section in \cite{Hax}
and the microscopic calculations of \cite{Lan00} show that a possible
oscillation signal relies strongly on the knowledge of the spectral properties
of $^{208}Bi$. In fact, the CC reaction cross section induced by $\nu_e$
scales almost as the square of the electron energy and is particularly
sensitive to the detailed structure of the excitation spectrum
as was already pointed out for the case of $^{12}C$ \cite{Vol}.
It is then important either to get 
the cross sections directly from the experiment
or/and to obtain different theoretical estimates in order to know the
theoretical uncertainties and how they
affect the reaction cross sections. 
This is
crucial when  the  impinging neutrino energy increases 
because not only the allowed Gamow-Teller (GT) and Isobaric Analogue
State (IAS) contribute significantly to these cross sections
but also forbidden
transitions, of first-, second-, third-order 
(which are not very well known experimentally).

In this paper, we present new theoretical results
for the CC $\nu_e + ^{208}Pb \rightarrow ^{208}Bi + e^-$ 
reaction cross section.
Our calculations, as opposed to \cite{Lan00},
are performed in a self-consistent
charge-exchange Random-Phase-Approximation (RPA) with effective Skyrme
forces. 
Contrary to all the previously published calculations,
we present  non flux-averaged 
cross sections, obtained for both low-energy  $\nu_e$ and
high-energy $\nu_{\mu}$.
These reaction cross sections given as a function of
neutrino energy span a large energy range.
They can be used to convolute with different neutrino fluxes
in various contexts, 
for example for future experiments with astronomical neutrinos
which are at present under study, for the very
recent terrestrial experiments such as the LSND ones \cite{Lan00}
to estimate the background,
or in the r-process nucleosynthesis.

We will emphasize the importance of 
the contribution of forbidden transitions and how it evolves as a function of
neutrino energy. This is often not taken into account in many 
present r-process nucleosynthesis calculations
and so the neutrino-nuclei cross sections are underestimated
(in \cite{surman} only the importance of first forbidden transitions
in neutron-rich nuclei was emphasized).

We will compare our results with presently available 
calculations \cite{Hax,Lan00}.
With this aim, we will present two different flux-averaged cross
sections, where
the neutrino fluxes are given by either
the DAR of $\mu^+$
and Decay-In-Flight (DIF) of $\pi^+$;
or by a Fermi-Dirac spectrum for a supernovae explosion.
Finally, we will discuss our results in relation to the suggested possible
oscillation signals that would use the spectroscopic properties of $^{208}Bi$.

The general expression for the differential cross section
as a function of the incident neutrino energy $E_{\nu}$ 
for the reaction 
$\nu_{l} + ^{208}Pb \rightarrow l + ^{208}Bi$ ($l=e,~\mu$) is 
\cite{Kubo}
\begin{equation}
\sigma(E_{\nu})={G^{2} \over {2 \pi}}cos^{2}\theta_C\sum_{f}p_lE_l
\int_{-1}^{1}d(cos \, \theta)M_{\beta},
\label{e:1}
\end{equation}
where 
$G \,cos \, \theta_C$ is the weak coupling constant, $\theta$ is
the angle between the directions of the incident neutrino and the outgoing
lepton, $E_l=E_{\nu}-E_{fi}$ ($p_{l}$)
is the outgoing lepton
energy (momentum), $E_{fi}$ being the energy transferred to the nucleus,
$M_{\beta}$ are the nuclear Gamow-Teller 
and Fermi type transition probabilities 
\cite{Kubo}.

In a nucleus as heavy as $Pb$ the distortion
of the outgoing lepton wavefunction due to the Coulomb field
of the daughter nucleus becomes large and affects
the integrated cross section considerably.
In our treatment of this effect we follow the findings
of ref.\cite{Engel}.
In ref.\cite{Engel} it is found that
the ``Effective Momentum Approximation'' (EMA)
works well
for high energy neutrinos.
This approximation consists in using
 an effective
momentum $p_l^{eff}=\sqrt{E^2_{eff}-m^2}$ where $E_{eff}=E-V_C(0)$
($V_C(0)$ is the Coulomb potential at the origin)
in calculating the angle integrated cross section
and multiplying eq.(\ref{e:1})
by $(p_l^{eff}/p_l)^2$.
It is also shown that the Modified EMA (MEMA) works better than EMA
for $\nu_{\mu}$ of low and high energies. 
In this approximation eq.(\ref{e:1})
is multiplied by $p_l^{eff}E_{eff}/p_lE_l$.
We use therefore this method 
in all our calculations of the $({\nu}_{\mu},{\mu}^-)$
cross sections. In the case of the $(\nu_{e},e^-)$ process,
the situation is somewhat more complicated.
The Fermi function works only for very low energies, 
namely $E_{e} \le 10~MeV$
(where $p_eR \ll 1$, $R$ is the nuclear radius),
whereas the EMA seems to be a good approximation for most energies
of the outgoing electrons \cite{Engel}. 
As in ref.\cite{Lan00},
for $\nu_e$ \cite{Lan01}, we treat Coulomb corrections by interpolating 
between the Fermi function at low electron energies
and the EMA approximation at high lepton energies.

To get flux-averaged cross sections it is necessary to
convolute (\ref{e:1}) by the neutrino flux $f(E_{\nu})$,
that is
\begin{equation}\label{e:2}
\langle \sigma \rangle_f = \int_{E_0}^{\infty} dE_{\nu} \sigma(E_{\nu}) f(E_{\nu}),
\end{equation}
$E_{0}$ being the threshold energy.
The choice of $f(E_{\nu})$ depends on the 
neutrino source and can be taken for example equal to the 
supernova neutrino energy spectrum given by transport codes
or the neutrino fluxes produced by a beam dump.

The nuclear structure model used to evaluate the transition probabilities
$M_{\beta}$ in (\ref{e:1}) is the charge-exchange Random-Phase-Approximation
(RPA). The  details of the approach can be found
in \cite{Col94}. The calculations we present
have been obtained in a self-consistent approach:
the HF  single-particles energies and wavefunctions as well as the
residual particle-hole interaction are derived from the same effective
forces, namely the SIII \cite{Bei75} and
SGII \cite{Gia81} Skyrme forces.
We have found that the model configuration space used is large
enough for the Ikeda and Fermi sum rule to be satisfied as
well as the
non energy-weigthed and energy-weighted sum rules
for the forbidden transitions \cite{pnRPA}.
The GT strength distribution we have obtained is
peaked at $19.2~MeV$, in agreement with the experimental value.
This main peak exhausts about $60 \%$ of the Ikeda sum rule.
The IAS results at $18.4 ~MeV$ and this value compares again well with
the experimental finding ($18.8~MeV$). 
Apart from these two resonances and the spin-dipole, 
the experimental knowledge about states of higher
multipolarity is rather poor. The recent experiment of Ref.
\cite{Zegers} shows that isovector monopole strength exists
in $^{208}Bi$ between $30$ and $45~MeV$ and 
in the present
calculation we find some strength in the same energy region.

In figs.1 and 2 we show the non flux-averaged
$^{208}Pb(\nu_{e},e^-)^{208}Bi$ and 
 $^{208}Pb(\nu_{\mu},\mu^-)^{208}Bi$ inclusive cross
sections 
as a function of the neutrino energy, for a mesh of energies, namely
$\Delta E=2.5~MeV$  for $E_{\nu_{e}}$ and
$\Delta E=5.0~MeV$ for $E_{\nu_{\mu}}$.
The dashed line in fig.1 shows the cross section obtained when
only the Fermi function is used to include the Coulomb corrections.
The results shown have been obtained with the SIII force, but we
have found that with the SGII force we get quite similar
results.
All the multipolarities with $J \le 6$ are included.
We have checked that the contribution coming from $J = 7$
is small.
(Note that, for higher multipolarities, a mean field description,
neglecting the particle-hole residual interaction, can be used
to evaluate the transition probabilities (\ref{e:1})).
In the calculations we present the axial vector coupling constant
has been taken equal to 1.26.
Note that the use of an effective $g_a$
 to take into account the problem of the ``missing'' GT strength
will reduce the reaction cross section by $10-15 \%$
as it was already  discussed in \cite{Vol}.

Figure 3 shows the contribution of the different multipolarities
 to the
total cross section (fig.1),
for the impinging neutrino energies
$E_{\nu_e}=15, 30, 50~MeV$, which are characteristic
average energies
for supernova neutrinos. 
When $E_{\nu_e}=15~MeV$ (fig.3, up), $\sigma_{\nu_e}$ is dominated by the
allowed Gamow-Teller ($J^{\pi}=1^+$) transition. 
As the neutrino energy increases (fig.3, middle), the allowed IAS and
other forbidden transitions 
start to contribute significantly.
Finally, when $E_{\nu_e}=50~MeV$ (fig.3, bottom),
the GT and IAS transitions are
not dominating at all,  
the cross section is being spread over 
many multipolarities. 
These results suggest that  
r-process nucleosynthesis calculations such as \cite{Bor},
which include neutrino-nuclei reactions, should 
take into account forbidden transitions.
This may be even more important
if ${\nu}_{\tau},{\nu}_{\mu} \rightarrow {\nu_e}$ oscillations occur, 
because in this case electron neutrino
may have a higher average energy than it is usually expected
from current supernovae models.

Let us now come to the comparison with other available calculations.
Table 1 shows our flux-averaged cross sections, 
in comparison
with those of
refs.\cite{Hax,Lan00}.
The low-energy neutrino flux is given by a Fermi-Dirac
spectrum \cite{Hax,Lan00}
\begin{equation}\label{e:3}
f(E_{\nu})={1 \over {c(\alpha)T^3}} {E_{\nu}^2 \over {exp\left[
      (E_{\nu}/T) - \alpha \right] +1}}
\end{equation}
where $T,\alpha$ are fitted to numerical spectra and $c(\alpha)$
normalizes the spectrum to unit flux. 
The values of the parameters $T$ and $\alpha$ have been chosen to
be able to compare our results with those of \cite{Hax,Lan00}.
As we can see from table 1, our predictions are in close agreement
(the difference is at most $20-30 \%$)
with \cite{Lan00}. The results of \cite{Lan00} have been obtained 
in a CRPA approach. 
A variation of $20-30 \%$ is actually to be expected for calculations
based on the same approach but using 
different parametrization (for example for single particle
wavefunctions and effective particle-hole interaction), because
of the sensitivity of  the flux-averaged cross sections to the 
detailed strength distributions \cite{Vol}, as we will discuss further.
On the contrary, our results and those of ref.\cite{Lan00} 
present significant differences
with those of \cite{Hax}, obtained using the allowed approximation
and including the IAS, the GT and the first-forbidden contributions
treated on the basis of the Goldhaber-Teller model.

We have checked that the differences do not come from
the higher order forbidden transitions  which are not
included in the calculations of \cite{Hax}.
The three calculations satisfy the same constraints, namely they
reproduce the centroid of the resonances and
satisfy the sum-rules.

We believe that the significant differences (by a factor of 2) 
with \cite{Hax}
may have two origins. 
The first possible origin might be the way the Coulomb 
corrections are treated.
In \cite{Hax}, the Coulomb
distortion
of the outgoing electron wave function was taken into account
by multiplying the cross section (\ref{e:1}) by a Fermi function.
In order to see the effect of using only the Fermi function instead 
of making an interpolation between the Fermi function and the
EMA approximation, we have calculated the 
reaction cross sections using these two possible corrections.
As figure 4 shows,
the two cross sections
have a quite different behaviour as a function
of the neutrino energy so that this difference on
the flux-averaged cross section
may vary according to the particular neutrino
flux considered. 
To get a quantitative idea of the variation, we have
calculated the flux-averaged cross sections 
by convoluting the two curves of fig.4
with (\ref{e:3}).
If we use the Fermi function only,
the reaction cross sections increase, on average, by $50 \%$.

The second possible origin of the discrepancies
between our work, \cite{Lan00} and \cite{Hax}
might be
the sensitivity of the flux-averaged cross sections to the 
detailed strength distributions in $^{208}Bi$. 
In fact, it has already been discussed in \cite{Vol},
that for low-energy neutrinos, the flux-averaged cross
sections are very sensitive to the energy of the excited
states in the final nucleus. The reason is twofold. 
First, due to the small electron mass, the non flux-averaged 
cross section (\ref{e:1}) scales as the square of energy
of the states. Second, the energy dependence of the neutrino
flux may emphasize differences in the non flux-averaged
cross sections due to variations in the energy of the states. 
As it was discussed in \cite{Vol}, these two effects may
modify the flux-averaged reaction cross sections by $20-30 \%$.

To complete our comparison with the calculations of \cite{Lan00},
we have calculated two more flux-averaged cross sections, using the 
neutrino fluxes
of both $\nu_{\mu}$ coming
from the DIF of $\pi^+$
and $\nu_{e}$ coming from the DAR of $\mu^+$.
The neutrino fluxes $f(E_{\nu})$ were taken from \cite{Imlay}.
These neutrino fluxes have been used in the recent experiments
$\nu_{\mu} \rightarrow \nu_{e}$ \cite{LSND1,KR0},
$\bar{\nu}_{\mu} \rightarrow \bar{\nu}_{e}$ \cite{LSND2,KR1}
or ${\nu}_{\mu} \rightarrow {\nu}_{x}$ \cite{KR2}
performed by the LSND and KARMEN collaborations.
The $DAR(\nu_{e},e^-)$ cross section
calculated 
is $\sigma_{DAR}=44.39 \cdot 10^{-40}~cm^2$ 
which is very close to 
$ 36.2 \cdot 10^{-40}~cm^2$ 
obtained in \cite{Lan00}.
On the contrary,
our $DIF(\nu_{\mu},{\mu}^-)$ is $\sigma_{DIF} = 399.2 \cdot 10^{-40}~cm^2$;
whereas the one of \cite{Lan00} is $115 \cdot 10^{-40}~cm^2$.
We believe that some of the disagreement may come from differences
in the strength distributions of the high order (higher than 2) 
forbidden transitions.
In fact, contrary to the reactions of neutrinos on light nuclei 
such as carbon, where these states  contribute
only
by $20 \%$ to the total DIF cross section, 
their contribution represents $65 \%$ of the total cross section
when the nucleus is  as heavy as
lead.

Let us finally discuss 
the two possible neutrino oscillation
$\nu_{\mu},\nu_{\tau} \rightarrow \nu_{e}$
signals based on the spectroscopic properties of $^{208}Bi$
excited in the CC reaction that have been proposed recently.
In \cite{Hax}, it was shown that the 2-neutron
events associated with the deexcitation of $^{208}Bi$
are very sensitive to the mean electron neutrino energy.
This signal relies on the fact that most of the IAS, GT
and first-forbidden
strength distributions are above the $2n$ emission
threshold ($14.98~MeV$) in $^{208}Bi$.
Our results show that not only the allowed and spin-dipole strengths
are above this threshold, but also a fraction of
the strength distributions
associated with other forbidden transitions (fig.3)
will contribute to the $2n$ decay.
All the arguments given in \cite{Hax}
are based on the statistical calculations of
$1n$ and $2n$ decays.
The direct $1n$ emission
represents about $50 \%$ of the total width in the case
of the IAS, and
$5-10 \%$ in the case of the GT \cite{Col94}. 

In \cite{Lan00}, it was pointed out that
the energy distribution of the neutrons in the $1n$
events should form a peak  at low energy,
more or less pronounced according to the occurence or absence
of oscillations. This peak comes from the GT
strength distribution at around $7.6~MeV$ which
is located above the $1n$ threshold emission at $6.9~MeV$.
Our GT distribution also shows a peak at around $7.5~MeV$.
We have checked that its 
location is not sensitive to the choice of the effective
forces used.
Still one should be careful about conclusions, because
predictions of 
different models about 
the energy location and strength
of that peak are at variance.

In summary, we have presented the non flux-averaged
$^{208}Pb(\nu_{e},e^-)^{208}Bi$ and
$^{208}Pb(\nu_{\mu},\mu^-)^{208}Bi$
reaction cross sections, calculated 
in a self-consistent charge-exchange
Random-Phase-Approximation with Skyrme effective forces.
These predictions can be employed for very different purposes, such as
for the interpretation of
the recent experiments on neutrino oscillations
performed by the LSND collaboration
(where reactions induced by neutrinos on lead
contribute significantly to the background)
and to evaluate the feasibility of future projects 
in which lead should be used as detector for  
supernova neutrinos.
We have emphasized that forbidden transitions
contribute significantly  to the
neutrino-nuclei reaction cross sections even at the
``astrophysical neutrino energies'' and they
should be included in present r-process nucleosynthesis
calculations.
We have discussed the present status on the theoretical
predictions on the reaction cross sections for
the $\nu_e$ having typical energies from present
models on supernovae.
If on one hand our calculations agree with those
of ref.\cite{Lan00}, which are also based on RPA; on the other
hand, they both significantly disagree with those of ref.\cite{Hax}.
We point out that the origin of the discrepancy might
be mainly the different treatment of Coulomb
corrections, but also the sensitivity of the reaction cross
sections to the detailed energy spectrum of the final
nucleus.
We have also compared our
flux-averaged reaction cross sections
with  $\nu_{\mu}$
coming from the
DIF of $\pi^+$
and with $\nu_{e}$ coming from the DAR of $\mu^+$,
with the ones of \cite{Lan00}.
As expected, the DAR cross sections are very close.
On the contrary our DIF cross section differs 
significantly from the one of  \cite{Lan00}.
We have pointed out that the two predictions may differ because
of differences in the strength distributions of 
forbidden transitions of high multipolarity which
represent the main contribution in reactions of neutrinos
on nuclei as heavy as lead.
Finally, we have discussed our results in relation with
recently proposed signals to measure 
 supernova neutrino oscillations.

\vspace{0.4cm}
This work was supported by the US-Israel Binational Science Foundation.

\begin{figure}
\begin{center}
\includegraphics[angle=-90.,scale=0.55]{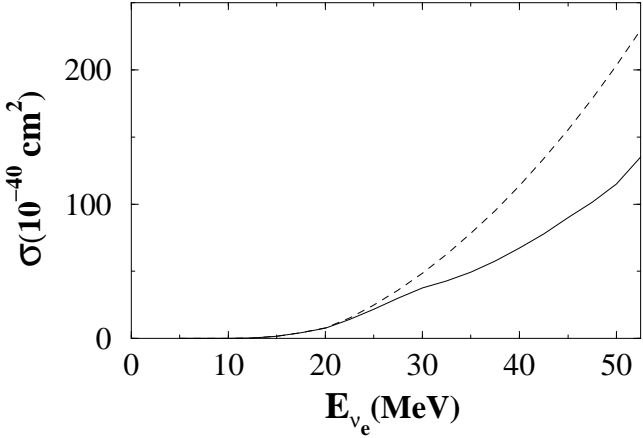}
\end{center}
\protect\caption{Differential $^{208}Pb(\nu_{e},e^-)^{208}Bi$ cross section as a function
of electron neutrino energy 
for a mesh of energies ($\Delta E_{\nu_e}=2.5 ~MeV$). 
As far as the treatment of the Coulomb distortion is concerned, 
the results are obtained by interpolating between
the Fermi function, good at low electron energies 
and the Modified Effective Momentum
Approximation, good at high outgoing electron energies.
The dashed line shows the result obtained if only a Fermi function
is used.}
\end{figure}

\begin{table}
\caption{Flux-averaged  cross sections ($10^{-40}~cm^2$) 
obtained by convoluting
the inclusive cross sections of fig.1 by a Fermi-Dirac spectrum (3)
for neutrinos emitted in a supernova explosion. 
Different temperatures $T$ and $\alpha$ values are considered.
The results of recent calculations are shown for comparison.}
\begin{center}
\begin{tabular}{lccc} 
$(T,\alpha)$  & this work & ref.\cite{Lan00} & ref.\cite{Hax} \\ \hline 
$(6,0)$ & 14.06 & 11. & 27.84 \\
$(8,0)$ & 25.3 & 25. & 57.99 \\
$(10,0)$ & 34.91 & 45. & 96.14 \\
$(6.26,3)$ & 25.21 & 21. & 47.50  
\label{tab:1}
\end{tabular}
\end{center}
\end{table}

\begin{figure}
\begin{center}
\includegraphics[angle=-90.,scale=0.3]{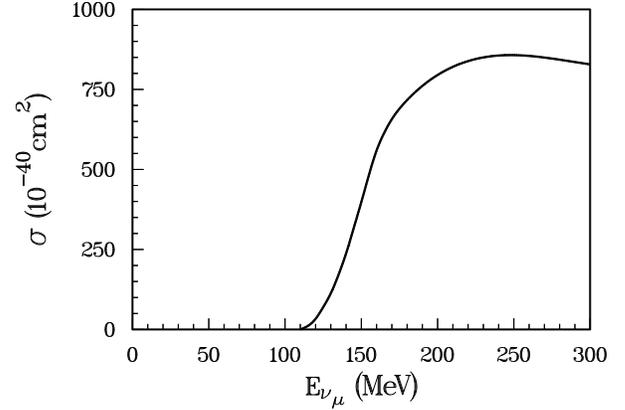}
\end{center}
\protect\caption{Differential $^{208}Pb(\nu_{\mu},\mu^-)^{208}Bi$ 
cross section, obtained with the MEMA approximation,  as a function
of muon neutrino energy for a mesh of energies 
($\Delta E_{\nu_{\mu}}=5.0 ~MeV$).}
\end{figure}

\begin{figure}
\begin{center}
\includegraphics[angle=-90.,scale=0.55]{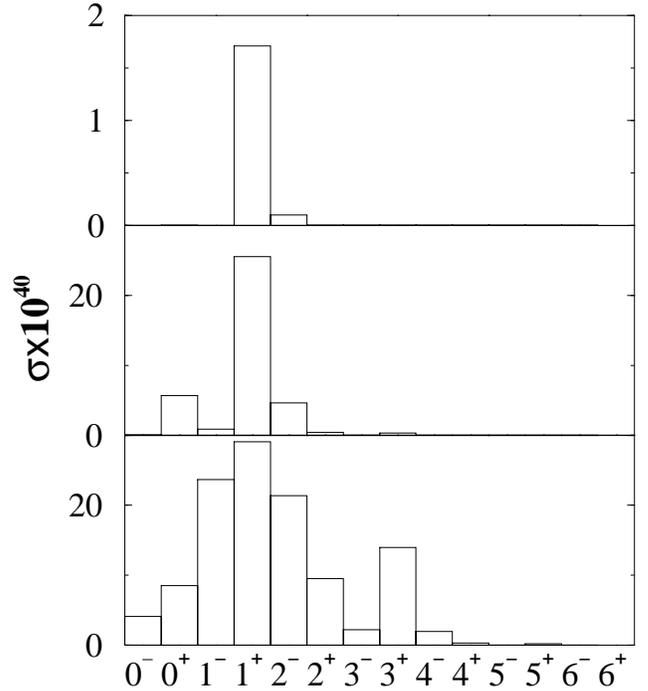}
\end{center}
\protect\caption{Contribution of the different multipolarities to
the differential $^{208}Pb(\nu_{e},e^-)^{208}Bi$ cross section $(10^{-40}~cm^2)$
of fig.1 for $E_{\nu_e}=15~MeV$ (up), $30~MeV$ (middle),
$50~MeV$ (bottom).}
\end{figure}

\end{document}